\documentclass{article}

\usepackage{arxiv}

\usepackage[utf8]{inputenc} 
\usepackage[T1]{fontenc}    
\usepackage{hyperref}       
\usepackage{url}            
\usepackage{booktabs}       
\usepackage{amsfonts}       
\usepackage{nicefrac}       
\usepackage{microtype}      
\usepackage{lipsum}		
\usepackage{graphicx}
\usepackage{natbib}
\usepackage{doi}

\title{Expansion and evolution of the R programming language}

\author{ \href{https://orcid.org/0000-0002-8550-2661}{\includegraphics[scale=0.06]{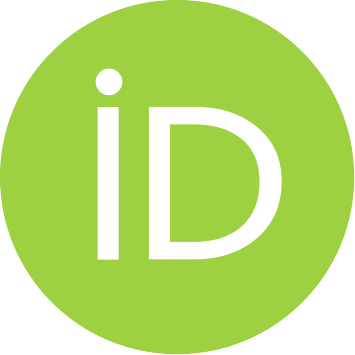}\hspace{1mm}Timothy L~Staples}\\
	School of Biological Sciences\\
	University of Queensland\\
	St Lucia, QLD, Australia, 4072 \\
	\texttt{timothy.staples@uqconnect.edu.au} \\
}

\renewcommand{\shorttitle}{R Evolution}

\hypersetup{
pdftitle={Expansion and evolution of the R programming language},
pdfsubject={cs.P},
pdfauthor={Timothy L Staples},
pdfkeywords={First keyword, Second keyword, More},
}

\begin{document}
\maketitle

\begin{abstract}
Change in language use is driven by cultural forces; it is unclear whether that extends to programming languages. They are designed to be used by humans, but interaction with computer hardware rather than a human audience may limit opportunities for evolution of the lexicon of used terms. I tested this in R, an open source, mature and commonly used programming language for statistical computing. In corpus of 360,321 GitHub repositories published between 2014 and 2021, I extracted 168,857,044 function calls to act as n-grams of the R language. Over the eight-year period, R rapidly diversified and underwent substantial lexical change, driven by increasing popularity of the tidyverse collection of community packages. My results provide evidence that users can influence the evolution of programming languages, with patterns that match those observed in natural languages and reflect genetic evolution. R’s evolution may have been driven by increased analytic complexity, driving new users to R, creating both selective pressure for an alternate lexicon and accompanying advective change. The speed and magnitude of this change may have flow-on consequences for the readability and continuity of analytic and scientific inquiries codified in R and similar languages.
\end{abstract}


\section{Introduction}

The reality of a language is not defined by the words that exist, but the words that are used \citep{Bochkarev2014}. Lexical evolution, the change in language use over time, is driven by a range of interacting social and cultural forces \citep{Bochkarev2014, Steels2005}. As well as stochastic drift \citep{Roberta2018, Newberry2017}, development of new domains and knowledge drives the formation of new words, increasing language size via the need for more specialized and specific terminology \citep{Ortlieb2022}. Language size is also restricted by cultural pressures to align word use \citep{Mark2019, Teich2021}, and the need for “optimal encoding” to communicate information efficiently \citep{Ortlieb2022}. Young languages rapidly expand, but this trend slows as languages mature and the need for new terms reduces, with speakers eventually converging on agreed language conventions \citep{Teich2021, Petersen2012}.\\

These forces are well described for natural languages, but it remains unclear whether they drive designed languages like programming languages. In some ways, software programs resemble books. Both are codified language that communicate meaning unidirectionally to an audience, eliciting a response \citep{Eden2007, Wang2014, Demey2018, Tylman2018}. In the case of programming languages, functions within programs execute algorithms, causing computer hardware to respond in predictable ways \citep{Gazoni2018}. So long as software applications remain constant, the human-computer dynamic of programming languages should make them resistant to the kind of linguistic change observed in natural languages. Despite this, software is still written and read by humans. Programming languages are subject to turnover, as new languages are designed, become popular, and eventually are superseded; most linguistic research has focused on this scale, making comparisons between programming languages \citep[e.g.,][]{Vasilescu2013}. But this turnover can also occur within programming languages, both through formal revisions and updates to their vocabulary and syntax and via user-created extensions.\\

Software may be subject to internal linguistic evolution like that of natural languages, but this hypothesis has been largely untested. Software linguistics, the application of linguistic tools and theories to software languages \citep{Falkoff1982}, has rarely been studied in earnest, specifically on whether, and how, the lexicon of software languages changes over time \citep{Favre2011}. The popularity of version control hosting sites such as GitHub provide access to a large and robust set of corpora to study the patterns of programming language evolution. Here I pair linguistic and ecological perspectives to understand the change in diversity and composition of programming functions as evidence of lexical evolution in the R language, accessing scripts housed in 360,321 GitHub repositories.\\

R is an ideal software language to test for evidence of lexical evolution. It is ranked in the top 20 most used programming languages \citep{TIOBE2022} and is freely available, creating a broad user base. It is a language for data analysis and statistical inference, restricting language use cases beyond those of more general-purpose programming languages; a broad variety of texts in corpora can limit the inferences that can be drawn \citep{Pechenick2015}. R has been available since 1993, creating a long legacy of training and use in the scientific and data analytic fields. It also has a history of package creation to extend and replace R’s functionality, which has been extremely popular with R users. This continued development cycle, alongside user-led package development, creates theoretical potential for R to evolve like a natural language, influenced by cultural forces from its user base. Finally, R has had a continual development cycle since its revision from S, without separation of major versions (such as seen in Python 2.7 and Python 3).\\

With these GitHub repositories, I asked the following questions:

\begin{enumerate}
\item Has the diversity and composition of R functions in use changed over time, both across and within GitHub repositories?
\item To what extent has function use in R been supplemented by community-led development of packages?
\item Are there trends in function and package usage that highlight potential mechanisms of R language evolution?
\end{enumerate}

\section{Methods}

\subsection{GitHub web scraping and data processing}

\subsubsection{Data acquisition}

I accessed repositories (``repos'') published on GitHub between 1st January 2014 and 31st December 2021 using the API. I identified all repos published each day (up to a maximum of 1,000) with the R language flag. I then extracted every ``.R'' script file in each repo, a total of 5,443,762 scripts housed in 701,956 repos, authored by 243,208 different GitHub users. Within each script I employed a function-extracting algorithm that captured text to the left of an open parenthesis (“(“), the syntax for function calls in R. In addition, I recorded instances of three common operators which act as functions with different syntax (``\%in\%'', ``[]'' and ``\%>\%''). These functions are equivalent to 1-gram verbs as used in corpus linguistics studies. This resulted in a dataset of total of 254,738,400 calls of 1,938,744 unique functions across all repos.\\

Next, I associated each function with its associated R package. This included the ``base'' set of packages that is included in an R installation, as well as many popular community packages that must be downloaded separately. To do this I matched function names in the dataset with the list of functions in each package. This process operated without error for any function name that was unique to a single package. Some function names (e.g., ``plot'', ``predict'') were repeated across different packages (with different versions for specific object classes or contexts). The actual package used in these cases was obscured, so I attempted to identify the most likely package source. Where the function shared a name with a base function, I associated the function with the relevant base package, as the function was likely an extension of the base function applied in a package-specific context. This process was then repeated for all functions that did not match a base function, first for the set of ``recommended'' R packages. For community packages, I conducted an iterative process, whereby I identified the most common functions not associated with a base package, added the most likely package including that function, updated the list and repeated for the next most common function with no associated package.

\subsubsection{Data processing}

I processed these raw data to better capture overarching trends in R function usage over time. First, I removed scripts from 138,872 repositories that were updated 12 months or more after their initial creation date. As I was accessing R scripts in their most recent GitHub version, these repositories were likely a poor representation of function use trends at the time the repository was created.\\

Second, I removed functions that were not associated with an R package: this was almost all functions (1,279,873: 98.80\% of the total unique functions). These non-package functions came from several sources. First, I did not include functions from rarely used packages. Second, R functions are objects and can be created locally within scripts or sourced from other scripts ad-hoc, without needing to be included in a published package. Finally, some functions identified may have been algorithm results from non-function parentheses, such as from mathematical operations or from script comments. Despite the number of non-package functions, they only represented 11.84\% of function calls in the dataset, owing to their rarity: 1,116,596 (91.15\%) of these non-package functions were called five or fewer times across the entire dataset.\\

I then filtered functions via a sampling threshold (Fig. S1). I retained only functions that were used in at least 0.1\% of repos in at least one given calendar month across the study period. This month-based test allowed for functions that were newly or historically popular to be included even if they fell below a sampling threshold averaged across the study period. A total of 1,188 functions exceeded the sampling threshold.\\

Finally, I removed 225,820 repos that only contained a single function. These were the equivalent of texts with a single word and contained insufficient data to estimate function diversity.\\

After processing, the dataset consisted of 3,267,885 scripts from 591,493 repositories created by 223,491 GitHub users. These repos contained a total of 143,316,514 calls of 1,188 functions from 133 packages.

\subsection{Function diversity over time}

To capture trends in function use across the study period, I used information theory metrics that capture different aspects of diversity \citep{Roswell2021,  Hill1973}, calculated both across all repos published each calendar month, providing an estimate of total R functions in use (``gamma diversity''), and within each repository (``alpha diversity''). Diversity estimates are inexplicably related to sensitivity to underlying relative abundance of different entities (functions in this case). I calculated gamma and alpha diversities with two different sensitivities, unified under the Hill number concept, using orders (q) of 0 and 1 \citep{Hill1973}. Hill order 0 is insensitive to underlying relative abundances: functions each contribute one count to Hill order 0 estimates regardless of how abundant they were, translating to a count of unique functions used. Hill order 1 is the exponent of Shannon’s Entropy (``perplexity'') \citep{Hill1973}, an estimate of the uncertainty in predicting the identity of an unknown n+1 function. Where only a few functions dominate (i.e., evenness of function use is low), there is low uncertainty, as the n+1 function is overwhelmingly likely to be a dominant function as well. Higher Hill order 1 scores equate to a diverse mix of evenly abundant functions.\\

To estimate these diversity metrics, I summed all function use in each repo, converting them to relative abundances that summed to one. This corrected function counts for differing repo code lengths. These relative abundances were used to calculate repo-specific Hill order 0 and 1 estimates to use in the alpha diversity models. For gamma diversity models, I averaged each function’s relative abundance across all repositories created during each calendar month.\\

Temporal trends in the four diversity variables (Hill 0 gamma, Hill 1 gamma, Hill 0 alpha, Hill 1 alpha) were modelled in separate Generalised Additive Models (gam function, mgcv package \citep{Wood2017}). Additive models allow for the fitting of flexible spline terms that can track non-linear trends through time. All models included one main fixed effect: number of months since January 2014 fit as a thin plate spline. Hill order 0 models with count responses were fit using a log link function and a negative binomial error distribution. Hill order 1 models were fit as Gaussian, with an identity link function. Alpha diversity models included natural log-transformed function count for each repo as a separate covariate to correct for sampling effects (a higher diversity of functions was expected in repos that used more functions).

\subsection{Functional composition}

I unpacked diversity trends by examining how the composition of R functions in use had changed over time, as well as temporal changes in the probability of function and package use. I visualised function composition change over time via a non-metric multidimensional scaling on monthly function relative abundances (those used for the gamma models) converted into a Bray-Curtis dissimilarity matrix, using the metaMDS function in the vegan package \citep{Oksanen2020}. This process constrains multidimensional differences in month-to-month function use into two dimensions (with lost information quantified via a metric of stress).\\

I also tested for how month-to-month variation in function composition was structured through time. To do this, I used a distance-based redundancy analysis \citep{McArdle2001, Legendre1999}. This partitions multivariate space into the portion that can be explained by a constraining variable, and the portion that cannot. I set the month of repository creation as a constraining time variable in this process, extracting the proportion of variation that was explained by time as a measure of the overall change in R function usage in the GitHub corpus.\\

I established function occurrence trends over time by using the presence or absence of each function in repos as a response variable in a binomial generalised linear model. Each calendar month was treated as a replicate, with the count of R repos from each month constituting a set of weighted trials. Repos containing a particular function were considered successes and vice versa. This was conducted separately for each function and modelled as a function of time from the end of the study period. I extracted the intercept (reflecting probability of function occurrence in December 2021) and slope (reflecting change in probability over time on logit scale) from each model. I then repeated these models, but aggregated functions into packages, treating the use of any package-specific function as a success, with separate models for each package.\\

I plotted intercepts and slopes of function and package change models against each other to establish patterns in the change in individual function and package use over the study period. These reflect word shift graphs used in linguistic research \citep{Gallagher2021}, except with fully quantitative occurrence probability (or “popularity”) on the y-axis instead of rank order. I separated functions and packages into groups of those included in a base R installation, the tidyverse package collection (including unofficial extension packages), and all remaining community packages.

\subsection{Trends in function categories}

I repeated the above diversity models and multidimensional scaling for each of the three function categories: base, tidyverse and other. Hill order 0 and 1 scores were re-calculated for each function category. Gamma diversity models were fit in the model structure described above, with separate splines and intercepts fit to each function category. As it was possible for a repository to contain zero functions of a particular category, the function category alpha diversity models were reconfigured as a hurdle process. I first modelled the probability of a repository including any functions for a category (binomial GAM, logit link function), followed by a model of relative abundance of each function category within each repository (with zero counts excluded: gamma GAM, inverse link function). Composition over time for each function category was conducted fitting using the functions from each function category in separate nMDS and distance-based redundancy analyses.

\subsection{Vignettes of change within synonymic function groups}

As examples of function change over time, I selected three groups of functions with similar use cases (``synonymic''). These included twelve functions that import tabular-style data (row and column data commonly entered and manipulated in spreadsheet-style software), seven functions that used index matching to combine tabular data objects together and nine functions that reshape existing data objects. These use cases are extremely common in data science, and the functions themselves were unique and not repeated across different packages.\\

I modelled the overall probability of each function group being used in a repo over the study period, using a binomial GAM, with a repo that used at least one of the functions constituting a success, using month of repository creation as a continuous thin plate spline predictor. I then repeated separate GAMs for each function within the group, using the same model format. I predicted trends over time from each of these models and captured the overall proportional change in both group and individual function probability by dividing the mean occurrence probability across 2021 by the respective mean probability across 2014.

\section{Results}

\subsection{Trends in the diversity of the R functional lexicon}

Over the study period, per-month diversity of R functions used across all repositories increased by \textit{c.} 30\% (Fig. 1A). Uptake of new functions increased primarily between 2014 and 2018, peaking in early 2020 before decreasing slightly (Fig. 1A). This trend was similar when weighting functions by commonness, with a larger peak at \textit{c.} 65\% more diverse than January 2014 followed by a \textit{c.} 20\% contraction during 2020 and 2021 (Fig. 1B). These broad trends of increasing diversity differed from patterns within repos (Fig. 1). Across the study period, the average number of functions used in a repo increased between 2015 and 2017, and then again from 2020, but peaked at 8 and 11\% on 2014 levels using Hill orders 0 and 1 respectively.\\

These patterns were not consistent across the four function categories. Base R functions were by far the largest category, containing 558 (46.97\%) of the functions that exceeded the sampling filter (Table S1). The diversity of these functions being used over time was largely stable, both across and within repos (Fig. 2). A greater diversity of functions from other packages were used across all repos as time progressed (Fig. 2A-B). This was reflected in an increased richness of functions from these packages in repos where they were used (Fig. 2D), it did not change the probability that a repo would include a function from these other packages (Fig. 2C). The most profound changes were in the tidyverse category: the overall diversity of tidyverse used increased by 1.6 times and 2.1 times on 2014 levels, for Hill order 0 and 1 respectively. (Fig. 2A-B). There was also an over fourfold increase in the probability of tidyverse functions being used in a repo (Fig. 2C-D).

\begin{figure}[h]
	\centering
	\includegraphics[]{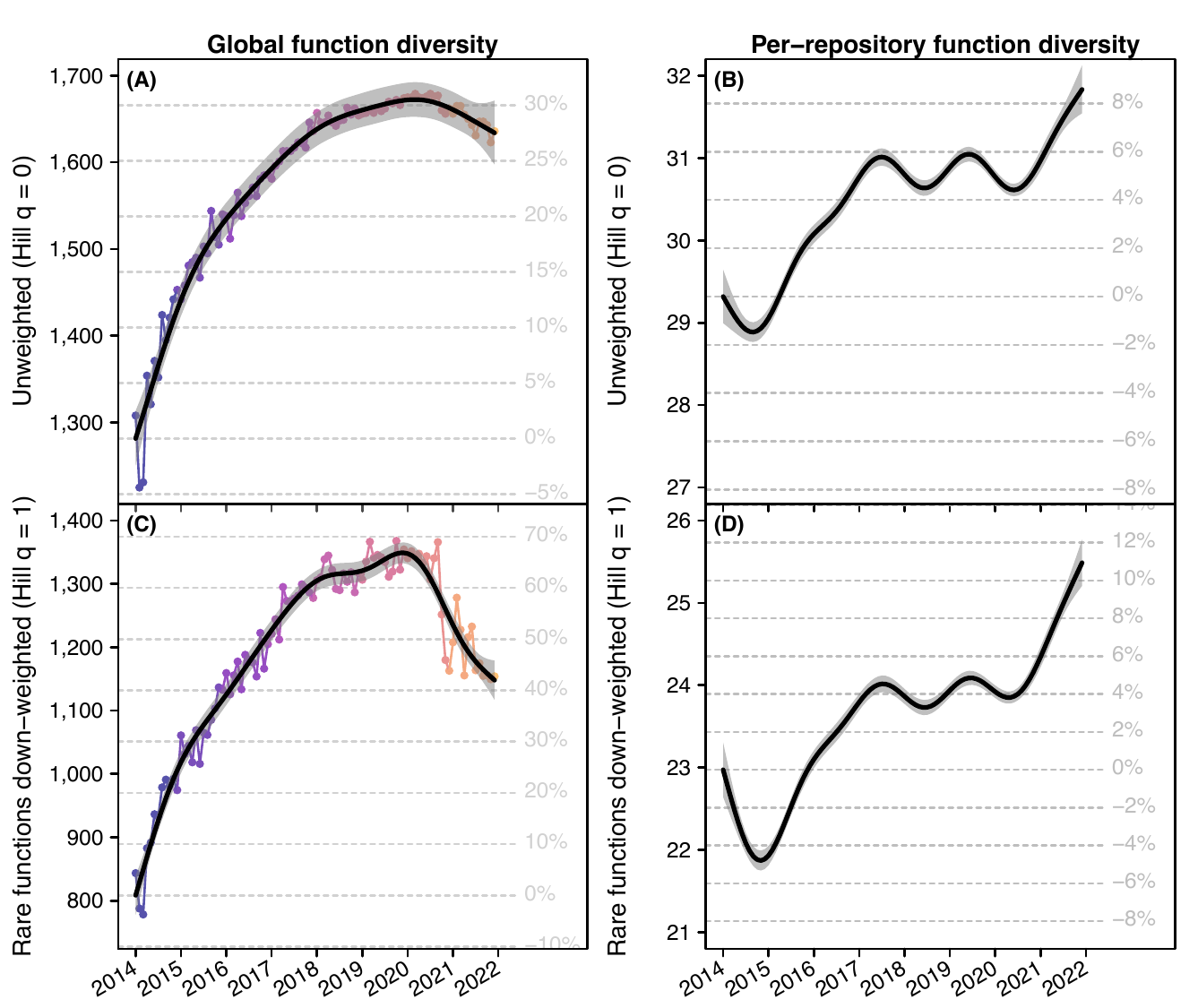}
	\caption{\textit{Diversity of R function lexicon over time, highlighting \textbf{(A)} average total monthly function diversity and \textbf{(C)} average function diversity within a single R repository. Black lines are mean estimates with accompanying 95\% confidence intervals. Grey dashed grid lines reflect proportional increases or decreases over time relative to January 2014 estimates. \textbf{(B)} and \textbf{(D)} are versions of \textbf{(A)} and \textbf{(C)} using a diversity metric that reduces the influence of rare functions and more closely represents the diversity of common functions.}}
	\label{fig:fig1}
\end{figure}

\begin{figure}[h]
	\centering
	\includegraphics[]{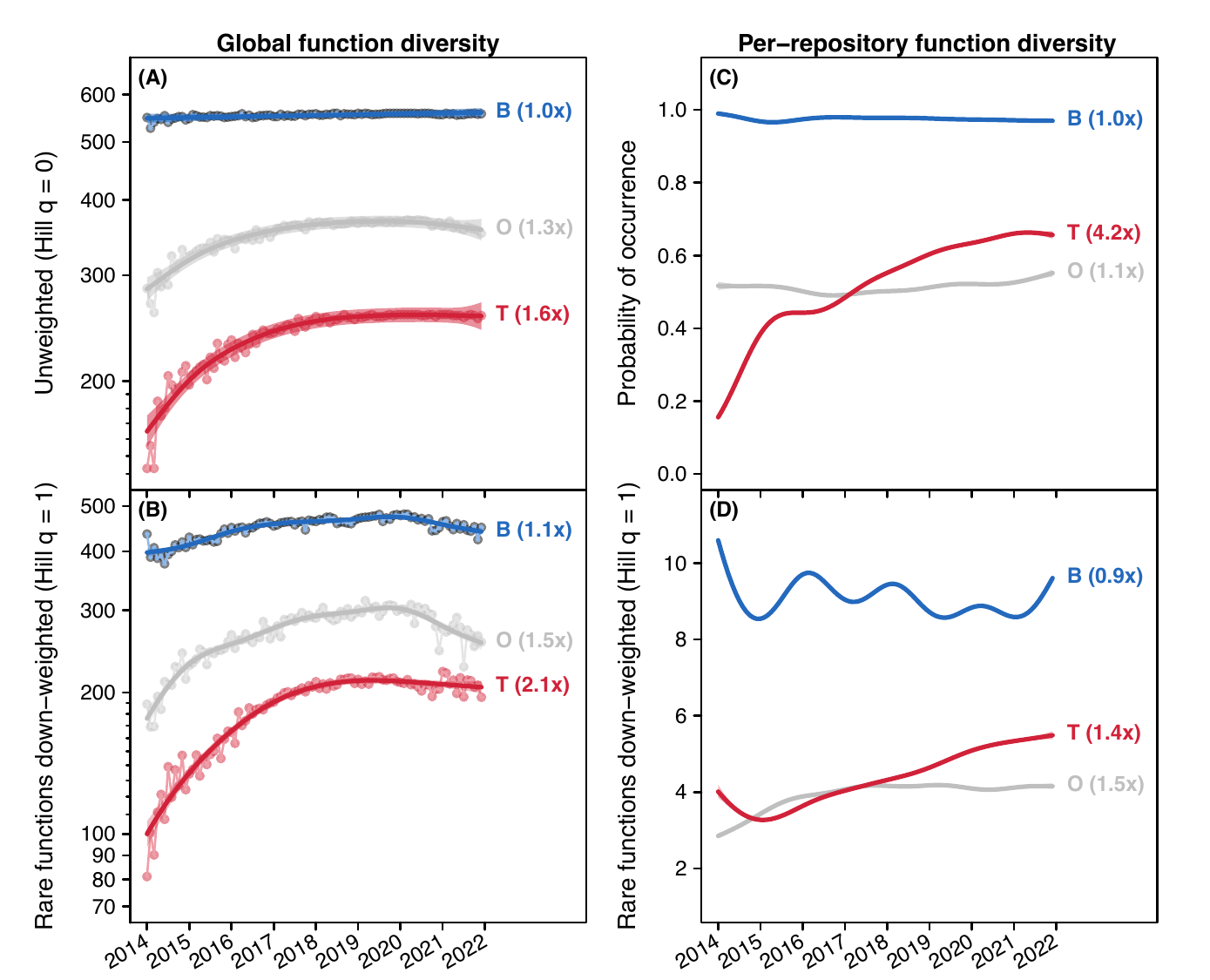}
	\caption{\textit{Diversity of R function lexicon over time, with functions split into three categories: functions included in the core R installation (``base'': B), those from a collection of interrelated community packages (``tidyverse'': T), and those from all other community packages (``other'': O). \textbf{(A)} Average total monthly function diversity for each function category. \textbf{(B)} is a version of (A) using a diversity metric that reflects common functions, downweighting the contribution of rare functions. \textbf{(C)} Per-repository probability of including a function from a given category. \textbf{(D)} Per-repository richness of functions in each category, including only repositories that contained a function from a given category (i.e., zeroes were excluded). Bracketed numbers following each trend reflect the proportional change in the metric for each function category between January 2014 and December 2021 (e.g., 2.0x = double).}}
	\label{fig:fig2}
\end{figure}

\subsection{Trends in function and package composition over time}

Time explained \textit{c.} 30\% of variation in function use in the distance-based redundancy analysis, a signal of strong temporal structuring in function composition (Fig. 3A). Patterns of temporal change were identified separately across all three function categories, each stronger individually than when all functions were considered together (Fig. S2). Variation in function composition was greatest in 2014, likely given small repository samples sizes, and compressed to discrete year-based clusters that moved across function composition space (Fig. 3A).

\begin{figure}[h]
	\centering
	 \begin{minipage}[c]{0.425\textwidth}
	\includegraphics[scale=0.775]{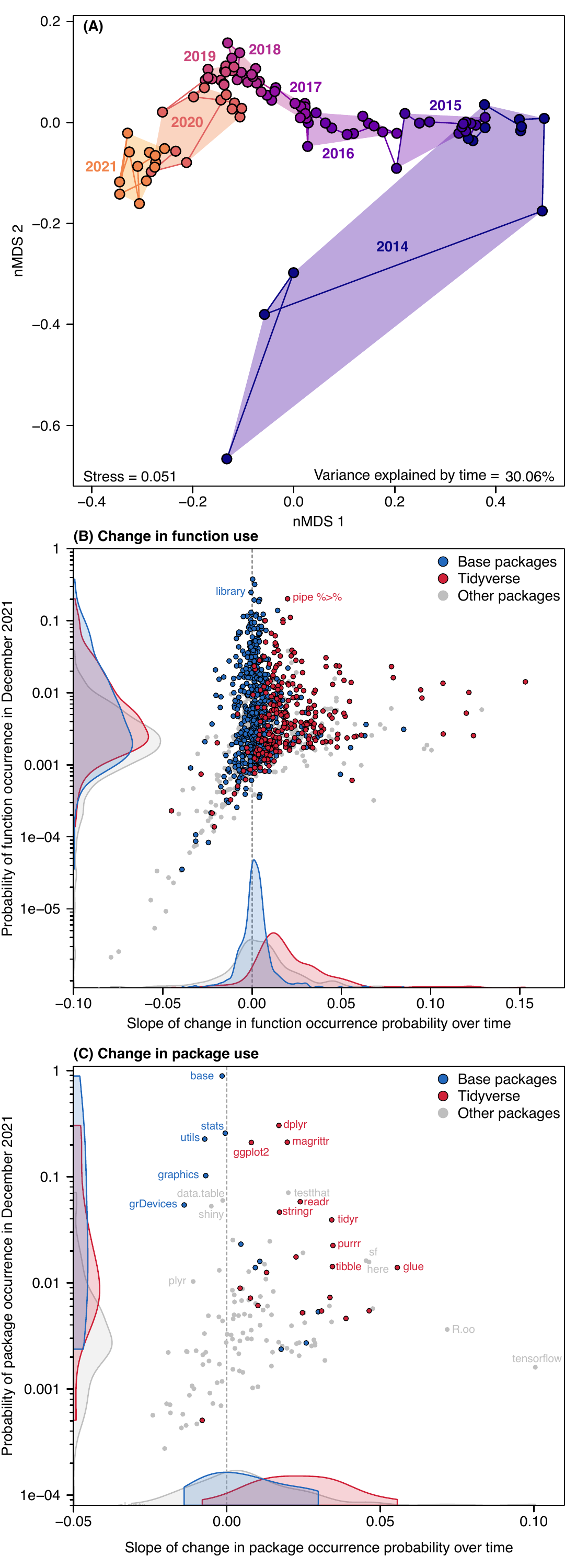}
	\end{minipage}\hfill
	 \begin{minipage}[c]{0.44\textwidth}
	\caption{\textit{Compositional change in R function usage over time. \textbf{(A)} non-metric multidimensional scaling of month-to-month function usage composition. Each month is represented by a point, with month dissimilarities in relative function abundance assessed via Bray-Curtis. Differences in function use were constrained to two dimensions, with lost information represented by stress. Month points are clustered in nMDS space based on relative function abundance, with points with similar function composition close together and vice versa. Calendar years are aggregated via coloured convex hulls. \textbf{(B)} Word shift graph showing function occurrence probability in December 2021 (y-axis), and the slope of change in this probability over time (x-axis). Point colour represents function category: functions from the base R installation (blue), the tidyverse collection of packages (red) and all other community packages (grey). Distribution of points from each function category are shown on x and y-axes via density curves. \textbf{(C)} As per (B), but where functions from packages were clustered together, with their occurrence probability modelled in aggregate.}}
	\end{minipage}
	\label{fig:fig3}
\end{figure}

\begin{figure}[htb]
	\centering
	\includegraphics[scale=0.65]{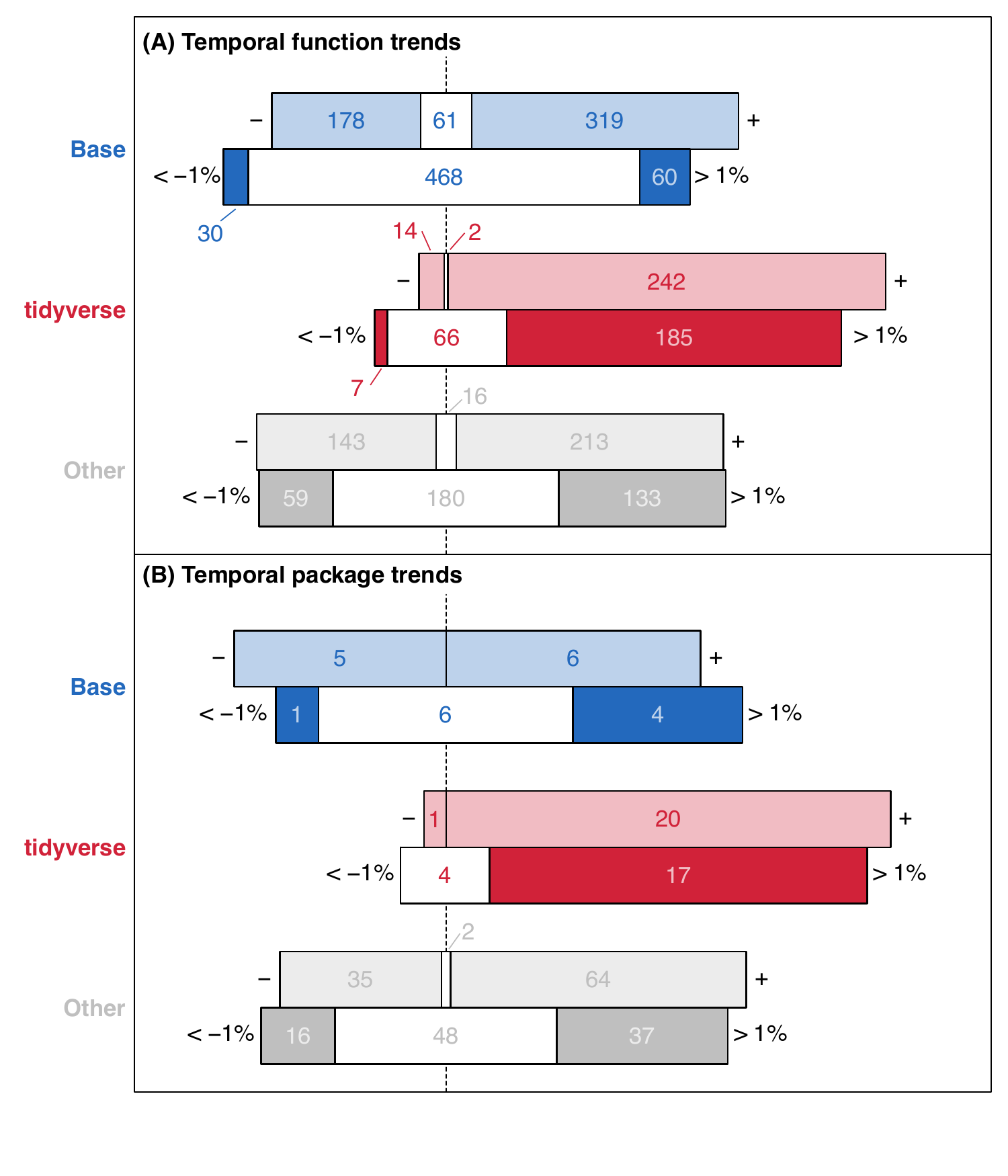}
	\caption{\textit{Change in R function and package occurrence probability between 2014 and 2021, in three categories: those included in the core R installation (``Base''), a collection of interrelated community packages (``tidyverse''), and all other community packages (``Other''). Functions in each category were split into those that decreased (left coloured boxes), increased (right coloured boxes) or were stable over time (white boxes), both significantly (top rows: p-value of slope of probability change over time < 0.05) and those that experienced a greater than 1\% relative month-on-month increase ( > 1\%'') or decrease (``< -1\%'') (bottom rows). Box widths are proportionate to the number of functions or packages in each category.}}
	\label{fig:fig4}
\end{figure}

Patterns of function occurrence followed a roughly log-normal distribution, with an over-abundance of rare functions (Fig. S4), fitting with Zipf’s law expected of word use within languages \citep{Piantadosi2014}. Twenty functions (1.68\%) were predicted to occur in more than 10\% of repositories in December 2021 (Fig. 3B). Sixteen of these top functions (80\%) were those included in the base R installation (Fig. 3B). Change in function occurrence over time was inversely related to function commonness; the statistical relationship between the occurrence probability and absolute change in probability over time was significant but weak (-0.179 $\pm$ 0.031, t = -5.763, p = 1.05e-8, R\textsuperscript{2} = 0.026).\\

Just over two-thirds of functions (774: 65.15\%) significantly increased in probability of use over time: 335 decreased (28.19\%) and 79 (6.64\%) showed no significant trend. Given large sample sizes in these models, significant trends did not necessarily translate into meaningful actual change in occurrence probability (i.e., effect sizes). Slopes presented as x-values in Fig. 3B were log odds ratios; their exponent were odds ratios, the proportional month-on-month change in occurrence probability. Taking a ``meaningful'' change as an odds ratio of 0.01 (i.e., a 1\% month-on-month change in relative occurrence probability), 378 functions meaningfully increased (31.82\%), 96 meaningfully decreased (8.08\%) and 714 did not meaningfully change (60.10\%).\\

Functions from base R packages were commonly used, and while most base R functions showed a significant increase or decrease in occurrence probability over time, only 90 (16.13\%) were strong, meaningful changes (Fig. 4). There was a slight trend towards base functions increasing in occurrence probability over time, with \textit{c.} 50\% more functions and packages increasing than decreasing.\\

Most tidyverse functions and packages increased over time (Fig. 3-4); only seven (2.71\%) tidyverse functions, and zero packages, decreased meaningfully over the study period (Fig. 4). All tidyverse packages bar one had a positive slope and some tidyverse packages (e.g., dplyr, magrittr, ggplot2) were as commonly used as base R packages by the end of 2021 (Fig. 3C).\\

Other community packages tended to be more rarely used and had a wider distribution of changes over time. Functions from other community packages were labile and more prone to change than base functions and packages, but with a similar two to one ratio of increases to decreases (Fig. 4). All bar one of the packages that decreased strongly in occurrence probability were from this other category (Fig. 3C, 4A).

\clearpage

\subsection{Vignettes of trends within groups of synonymic functions}

The three groups of synonymic functions exhibited similar overall declines in usage (Fig. 5A-C). In all three cases there was a dominant function that decreased substantially in usage over time (``read.table'', ``merge'' and ``melt'' respectively: Fig 5).\\

In functions that import external data, there appeared to be a shift towards R users preferring a format of input data rather than a particular function: the strongest trends were in functions that import comma-separated value files (.csv) or file formats from Excel (Fig. 5D). The base function ``read.csv'' maintained its occurrence probability across the study period (Fig. 5D). The base function that joined data objects together (``merge''), while still popular by 2021, was slowly being replaced by a collection of six tidyverse functions (Fig. 5E). The most commonly used reshape function (“melt”) is not included in base R, and base R functions with similar functionality were rarely used across the study period (Fig. 5F).\\

Tidyverse functions tended to be rarely used in 2014 but increased substantially in almost all cases. Across all synonymic groups, the occurrence probability of nine out of 13 tidyverse functions increased more than tenfold, up to a maximum of 77 times. Notably, no tidyverse functions decreased in occurrence probability (Fig. 5D-F). By contrast, the largest increase in a base functions and functions from other packages were ``stack'' and ``gather'', with a twofold and sixfold increase respectively (Fig. 5F).

\begin{figure}[!htb]
	\centering
	\includegraphics[scale=0.55]{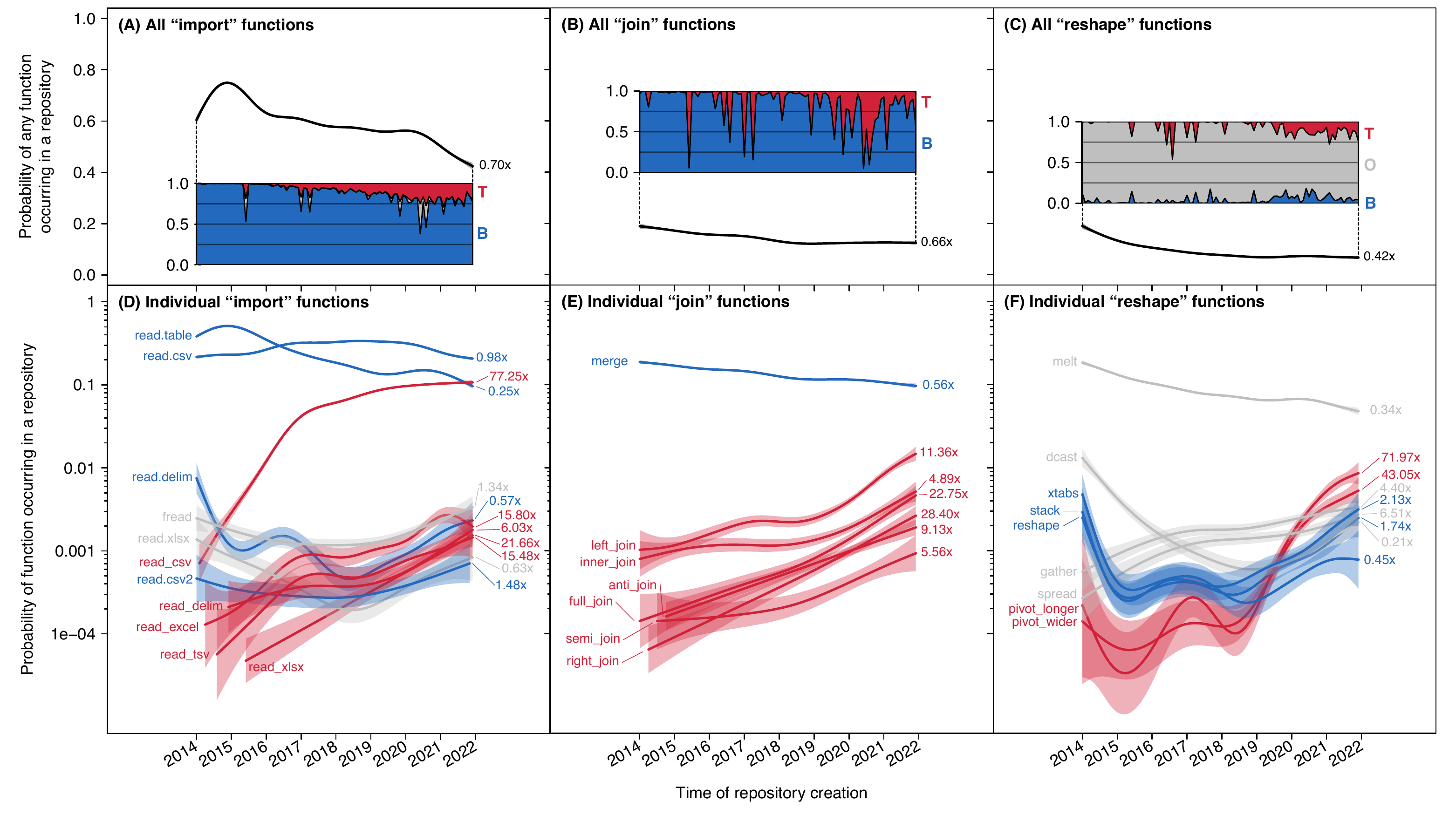}
	\caption{\textit{Occurrence probability patterns across (A-C) and within (D-F) groups of synonymic functions: those that (A, D) ``import'' external tabular data into R, (B, E) ``join'' tabular objects based on index matching and those that (C, F) ``reshape'' data objects. Y-axis values are the probability that a repository will use either any function (A-B) or each specific function (C-D). Functions are coloured based on their category (base, blue; tidyverse, red; other, grey). Right-hand numbers are proportional changes from 2014 to 2021 yearly means (i.e. 2.0x = doubling in occurrence probability). Area plots in (A-B) are proportional breakdown of all function calls over time by category: base (``B''), tidyverse (``T'') and other (``O'').}}
	\label{fig:fig5}
\end{figure}

\clearpage

\section{Discussion}

Over the last eight years, the functions being used in the R programming language has undergone rapid expansion and directional evolution, driven primarily by community-created addons. These patterns of lexical change are evidence that despite their designed nature, programming languages can change and evolve over time in a manner reflective of natural languages. Despite these similarities, observed lexical change in R has been rapid and extreme; some functions, especially those included in the popular tidyverse package collection, experienced more than tenfold increase in use in less than a decade, turning from rarities into core, commonly used programming verbs. The speed of language evolution in R may reflect rapid diversification expected in young languages but raises concerns about the readability of historical R programs by new users. Such rapid change may have flow-on consequences for the continuity of analytic and scientific inquiries codified in R and other programming languages with similar features. Overall, I provide evidence that online software repositories are robust, data-rich corpora to study the dynamics of young emerging languages, with potential to act as testing beds for linguistic hypotheses about language evolution and change.\\

\subsection{Equating programming languages to natural languages}

Even though programming languages are designed to interface with computer hardware, and their status as true `languages' is debated \citep{Schubert1989, Rao2009, Smaha2015}, they are designed by people and used by people \citep{Wing2006}. Given a desire for simple execution and comprehension, there is a drive to improve programming languages to make them more efficient and intuitive, aligning more closely with how humans solve problems \citep{Myers2002}. This occurs at the inter-language scale, where programming languages such as Julia were designed to fulfill specific limitations within scientific computing \citep{Nazarathy2021}. My results suggest this improvement happens within languages too. The third most used R function by the end of 2021 was ``library'', which loads an external package, reflecting the strong reliance of R users on non-core functionality. Most community packages in my analyses were rarely used across the entire study period. But in eight years tidyverse functions and packages rose from virtual obscurity to, in some cases, co-dominance. Rather than being rare outliers in a distribution of changing function usage, these patterns are markers of a broader paradigm shift in the building blocks that R programmers are using to solve problems.\\

The tidyverse brings a unique dialect of functionalization to R, providing a number of discrete functions to replace a single function that could accomplish a range of related tasks. The synonymic join functions outlined in Figure 5B showcase this. Base R’s ``merge'' function offers all the functionality of the six tidyverse functions via the specification of internal arguments. Base R packages tend to provide broad, flexible tools are modified internally to achieve specific outcomes, while the tidyverse instead provides a set of functions that each accomplish one goal. In natural language terms, base R provides general verbs, such as ``run'', and allows them to be modified by adjectives (e.g., ``fast'', ``slow'' and ``relaxed''). Tidyverse instead provides a series of specialised verbs (i.e., ``sprint'', ``jog'' and ``amble'').\\

This trend towards functionalization explains much of the observed increase in diversity of R functions. All increases in the diversity of R functions used came from non-base groups, and tidyverse exceeded the growth in all other community packages, both in overall count and when adjusted for function rarity, especially function richness within repositories. Not only were repositories over four times more likely to use tidyverse functions by the end of 2021, but GitHub repositories that used tidyverse used, on average, 25\% more functions than repositories with no tidyverse (Fig S2). Increasingly, R users appear to be drawing from a larger spectrum of more specific functional tools to accomplish their goals, driving rapid lexical evolution.\\

\subsection{Why is R evolving so rapidly?}

The rate of diversification and change in R functions appears faster than comparable rates observed in modern natural languages \citep{Petersen2012B, Baptiste2011}. R is currently at the early stages of language evolution, where rapid turnover is expected, driven by both diversification and consolidation of new words \citep{Petersen2012}. We observed patterns of diversification in global and repository-level function diversity, as well as in individual functions, although the contraction in global function diversity when abundance was considered may be an early sign of consolidation and simplification. On balance it appears the R language is still diversifying and increasing in size, potentially driven by cultural selection, a lag between function gains and losses (i.e., an extinction debt), or by features of R users that make the language more susceptible to stochastic lexical drift.\\

\subsubsection{Selection}

Big data operations with millions or even billions of observations are becoming more common, as are complex analytic processes such as quantitative genetic pipelines, machine learning, cloud computing and Bayesian statistical models. The size and complexity of tasks being performed in R has likely increased commensurately with increases in data size and availability, acting as selective drivers for the development of new functions to accomplish new goals efficiently. The lability of functions from other R packages may be an indication of this, as these packages comprise groups of specialised functions for specific data types (e.g., spatial data, genetic sequences) or analytic approaches (e.g., Bayesian modelling, machine learning). Over time old packages are likely to be superseded by either new analytic techniques or superior implementation of existing techniques.\\

The increasing complexity of data analyses may be incentivizing users of other statistical tools, such as spreadsheeting software, to learn R, increasing the population size of R speakers. Tidyverse provides an attractive entrance point for new R users, especially those with little mathematical, statistical, or programming expertise. The logic of tidyverse mirrors human sentence structure more closely than the mathematical logic of base R functions. Some evidence of this phenomenon could be the rapid increase in usage of ``read\_csv'' (Fig. 4D). This function is one of the defaults used by the popular RStudio graphical interface for R, which may be an indicator of the arrival of new R users.\\

The increase of new R users, driven by the need for more flexible and powerful statistical tools, may be driving a cultural, advective change in R away from some base functionality and towards the tidyverse \citep{Karjus2020}. Advection is the process by which some words become more popular not by direct selection, but by being associated with words that are being selected. As an example, the ``plyr'' and ``dplyr'' packages offer similar functionality, share developers and a design approach \citep{Wickham2022, Wickham2011}. The primary difference is dplyr’s inclusion and integration with the tidyverse. Over time dplyr’s use increased, while plyr’s use dropped (Fig. 3C). These trends need to be interpreted with caution, as some function names are shared between the two packages and were assigned to dplyr in my analyses, and observational data cannot uncover mechanisms. Despite this, patterns in a selection of non-overlapping functions from the two packages reflected the broader trends; plyr functions decreased while dplyr functions increased in use (Fig. S5).

\subsubsection{Extinction debts}

The rapid increases in overall R function diversity may reflect extinction debts observed in ecological systems, where the actual loss of species can be delayed for decades beyond the change in conditions that is driving extinction \citep{Tilman1994}. This pattern has been observed as a lag between initial dilation and subsequent contraction of natural languages \citep{Petersen2012}, particularly those requiring specialized terms like scientific English \citep{Ortlieb2022}. Almost no R functions that passed the sampling filter went extinct across the study period, although some (such as ``read.table'' and ``merge'' in Fig. 5) reduced substantially in usage, and function diversity weighted by abundance decreased from 2020. To project these decreases into the future, R function diversity may, or is beginning to, contract as older users are removed from the population, replaced by new users more likely to preferentially use the more popular functions.\\

Unlike the extinction of words, which only pose issues for historical scholars and translators, to maintain backwards compatibility of software programs, languages need to be able to understand and process all functions, even those no longer in use. This may lead to bloated installation size and documentation. As an example, there were 1,997 base R functions used in repositories that were excluded by the sampling filter: 3.57 times more than were included. While this is unlikely to be a problem currently, it may escalate in the future. Of the 18,402 packages listed on R’s official package repository, at time of writing, 6,293 (34.20\%) had not been updated since R version 4.0 released in April 2020. Some of these may be orphaned, with no active developers, and could have incompatibilities with changes to base-level R or with other interlinked packages that have been updated more recently. R programs written using these packages may be non-functional, or even incomprehensible to readers without prior knowledge of the functions in abandoned packages.

\subsubsection{Stochastic drift}

Population sizes for programming languages are small; they are learned only for specialized purposes, and direct communication is between a user and a computer. This may make them more susceptible to random drift, like genetic evolution patterns in small populations \citep{Ellstrand1993}. R’s design use, scientific and statistical computing, may be especially vulnerable to communicative rifts between users, as programs are often written by individuals or collaborations of speakers in small corporate or research silos, limiting the cross-population flow of cultural forces. The generation time of R users is also likely much shorter than natural languages, which is tied to lifespan. Instead, turnover is linked to time spent in related industries, which may be extend from an entire working life to the length of a PhD.\\

While a large corpus, GitHub is also likely a biased subset of the R-speaking population. Many R programs are never published publicly, and GitHub users may reflect a ``power user'' subpopulation more willing to adapt and change their language usage, as well as develop new R packages. This subpopulation may be more susceptible to cultural forces that drive trends in function use. Many of the GitHub users in the dataset only published a single repository, likely increasing the influence of inter-user variance on observed diversification. Despite this, there are similarities between published R repositories and published authorial works in natural languages: corpora of published books underpin many studies of language evolution \citep[e.g.,][]{Baptiste2011}.

\subsubsection{Functions as language evolution indicators}

My analysis focused specifically on functions and excludes all other types of R objects, which may have amplified the signal of lexical change over time. Notably, I excluded the names of non-function objects, which resemble nouns in natural languages. Focusing solely on verbs may have amplified the signal of language evolution, as noun conventions could both increase lexical variation (e.g., user variation in object naming conventions) as well as decreasing turnover (e.g., use of common object names like “data”, “a”, “x”). 

There may also be features of R that are more resistant to change, like grammatical rules and structures such as word order in natural languages \citep{Nichols1992, Hubler2022}. One such grammatical shift is the transition towards “piping”, linking R statements together via programmatic semicolons. This fundamentally shifts the grammar of R as a language, even if it does not alter functions. Pipes were initially part of the tidyverse (and marked as such in my analysis) but were included in base R from 2020. By the end of 2021 pipes were observed in 20\% of R repositories, equal fourth for most popular function (Fig. 3B), a \textit{c.} 12-fold increase on 2014 usage. Despite this example of labile grammatical change, I targeted an aspect of R grammar to identify functions during data collection, that function calls are always paired with parentheses. This grammar has not changed over time, and there currently no competing methodology for this approach. In all there does not appear to be an innate resistance of programming grammar to change, so long as alternative structures are feasible and available to be selected by speakers.

\subsection{Conclusion}

Programming languages are a way to express a mental plan for a computer to implement. While programmers do not need to design language around the imperfect interpretation of a human reader, they are ultimately still humans writing language. The rapid transition in R towards tidyverse suggests that while R’s core design may have been internally consistent, it is being selected against by the population of users, accommodated by the easy publication and incorporation of community-written packages. This accords with broader preferences in programming languages, such as Python, which is popular for its ease-of-use and intuitiveness. Programming language designers should take human language preferences seriously, both in fundamental design, but also in allowing for natural progression of the language over time. R's design accomplishes this with its stated role as an ``environment'' for developing and implementing new data analysis methods, rather than a rigid set of programming tools \citep{R2022}. While rapid changes in R raise concerns about the continuity and permanence of programs written over time, allowing programming languages to evolve naturally, such as via community-developed packages, offers a way to crowd-source language improvements, leveraging the natural tendency of human speakers to converge and refine language use \citep{Ortlieb2022, Teich2021}.

\section{Acknowledgements}
I am grateful to Catherine Bowler, Natalie Jones and Victoria Reynolds for initial helpful discussions around study design and structure. I also thank the R Core Team responsible for managing the R Project for Statistical Computing, the many package contributors and anyone committed to the principles of Free and Open Source Software.

\section{Funding}
This work was funded by an Australian Research Council Discovery Grant (DP210100804).

\section{Data accessibility}
R code to access GitHub repositories, extract R scripts and function calls, as well as data and code to reproduce all analyses, figures, and tables, will be lodged with Zenodo and cited via a DOI number here in a subsequent preprint version.

\bibliographystyle{unsrtnat}
\bibliography{references}

\clearpage
\section{Supplementary material}

\setcounter{figure}{0}
\setcounter{table}{0}
\makeatletter 
\renewcommand{\thefigure}{S\@arabic\c@figure}
\renewcommand{\thetable}{S\arabic{table}}
\makeatother

\begin{table}[h]
	\caption{\textit{Count of R functions in major categories before and after sampling filter was applied. Functions were only retained if they occurred in at least 0.1\% of sample repositories in at least one calendar month between January 2014 and December 2021.}}
	\centering
	\begin{tabular}{lcccc}
		\toprule
       & Base functions & Tidyverse & Other packages & Total\\
		\midrule
Total count &  2,555 & 1,805 & 11,288 & 15,648\\
Above sampling filter & 558 & 258 & 372 & 1,188\\
\% Retained & 21.84\% & 14.29\% & 3.30\% & 7.59\% \\
		\bottomrule
	\end{tabular}
	\label{tab:table}
\end{table}

\begin{figure}[h]
	\centering
	\includegraphics[]{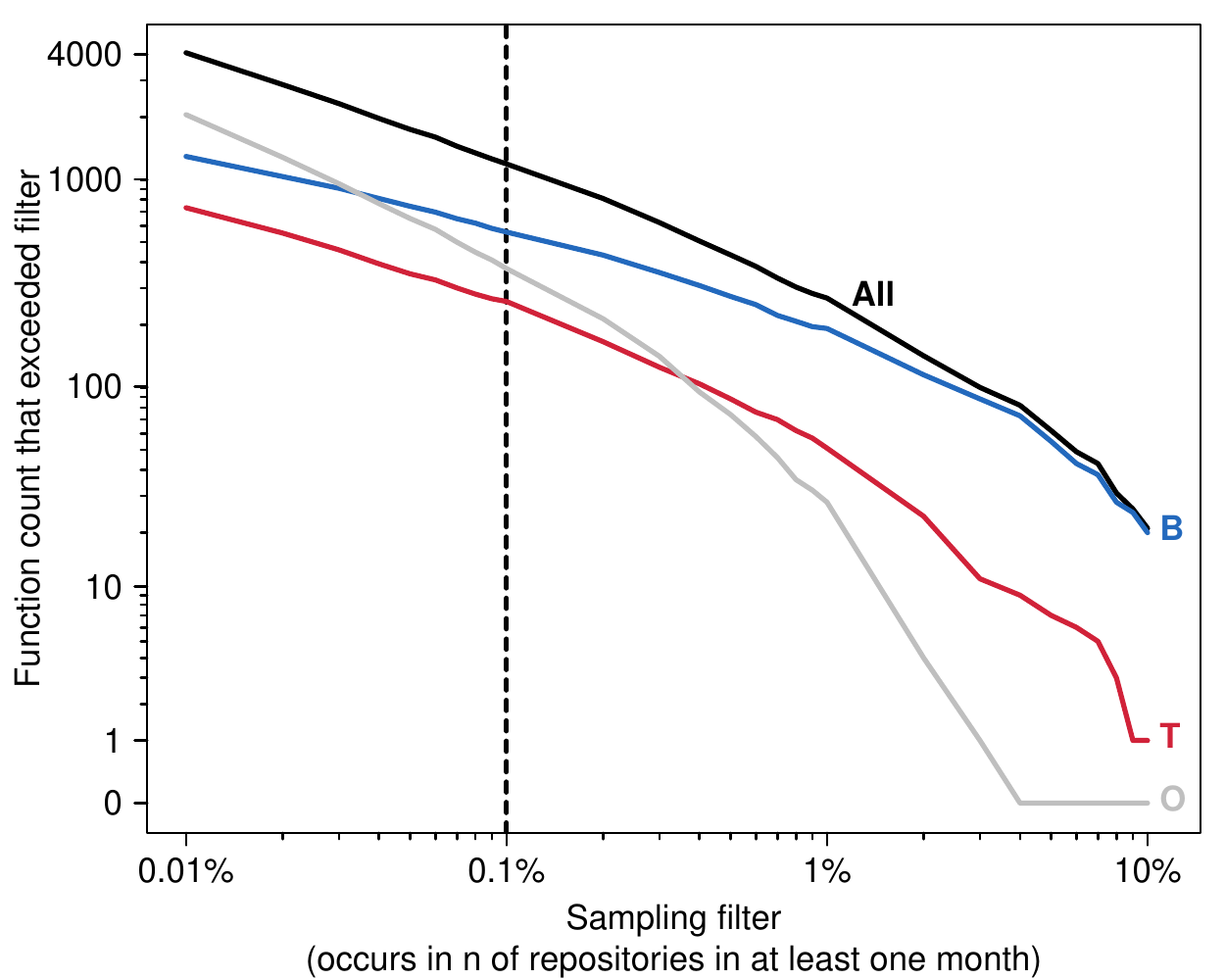}
	\caption{\textit{Sampling threshold sensitivity test. Number of R functions across all three major categories (``All''), as well as those in the base R installation (``B''), the tidyverse package collection (``T'') and all other community packages (“O”) that exceed a sampling threshold. Functions exceeded the sampling filter if they were used in more than n\% of GitHub repositories in at least one calendar month across the 2014-2021 study period. The dashed line reflects 0.1\%, the sampling filter applied to data used in the main text.}}
	\label{fig:figS1}
\end{figure}

\begin{figure}[h]
	\centering
	 \begin{minipage}[c]{0.425\textwidth}
	\includegraphics[scale=0.925]{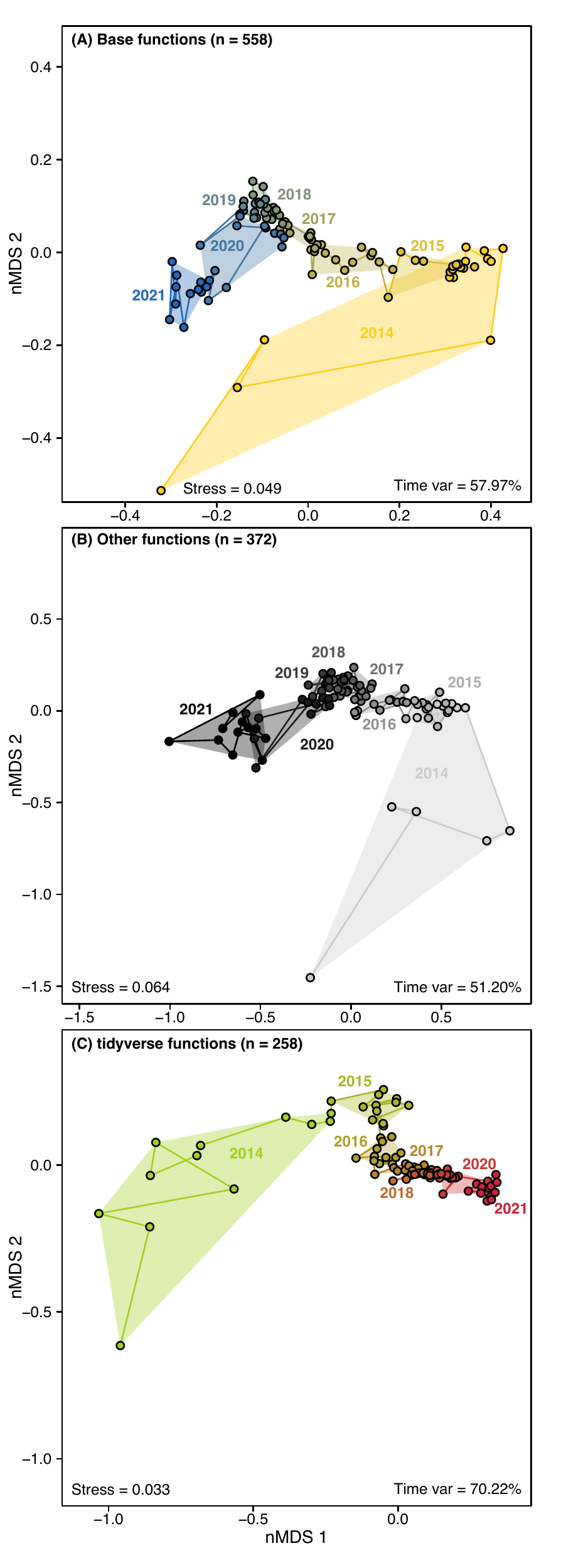}
	\end{minipage}\hfill
	 \begin{minipage}[c]{0.5\textwidth}
	\caption{\textit{Compositional change in R function usage over time in four function groups, represented as non-metric multidimensional scalings of month-to-month function usage composition. Each month is represented by a point, with month dissimilarities in relative function abundance assessed via Bray-Curtis. Differences in function use were constrained to two dimensions, with lost information represented by stress. Month points are clustered in nMDS space based on relative function abundance, with points with similar function composition close together and vice versa. Calendar years are aggregated via coloured convex hulls. Variance explained by time is expressed as a percentage of all variation in function composition, as determined by a distance-based redundancy analysis.}}
	\end{minipage}
	\label{fig:fig2}
\end{figure}

\begin{figure}[h]
	\centering
	\includegraphics[]{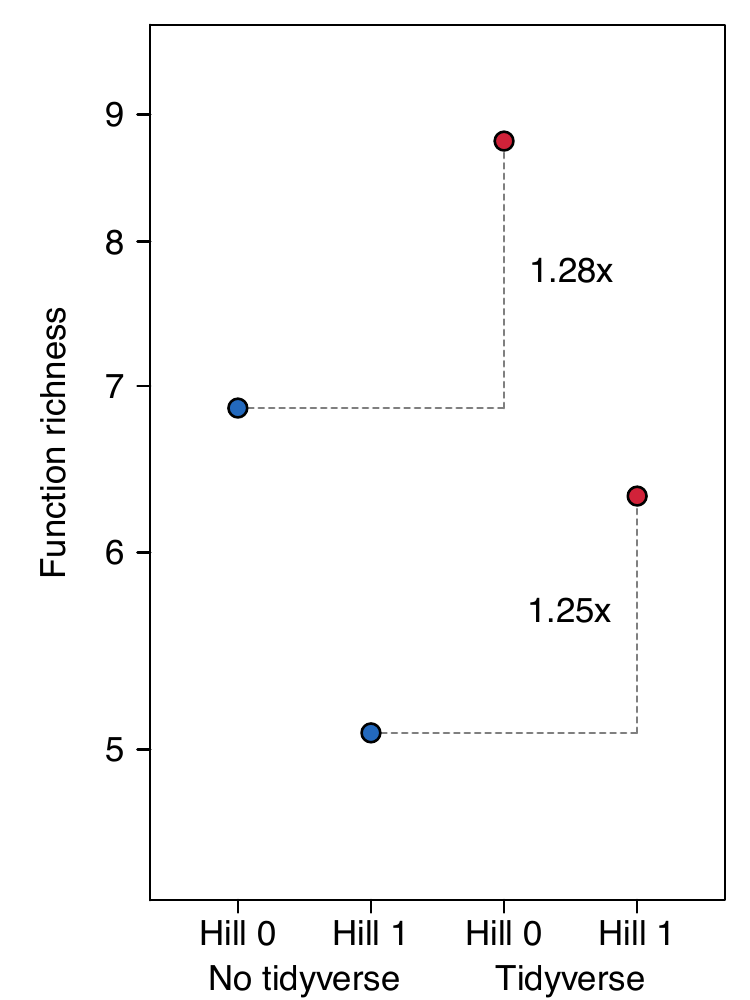}
	\caption{\textit{Estimated function richness in repositories that contained at least one function from a tidyverse package, and those that did not. Richness was estimated using Hill numbers orders 0 and 1; Hill order 0 is a count of unique functions, Hill order 1 is the exponent of Shannon’s entropy, and more closely reflects the richness of abundant functions.}}
	\label{fig:figS3}
\end{figure}

\begin{figure}[h]
	\centering
	\includegraphics[]{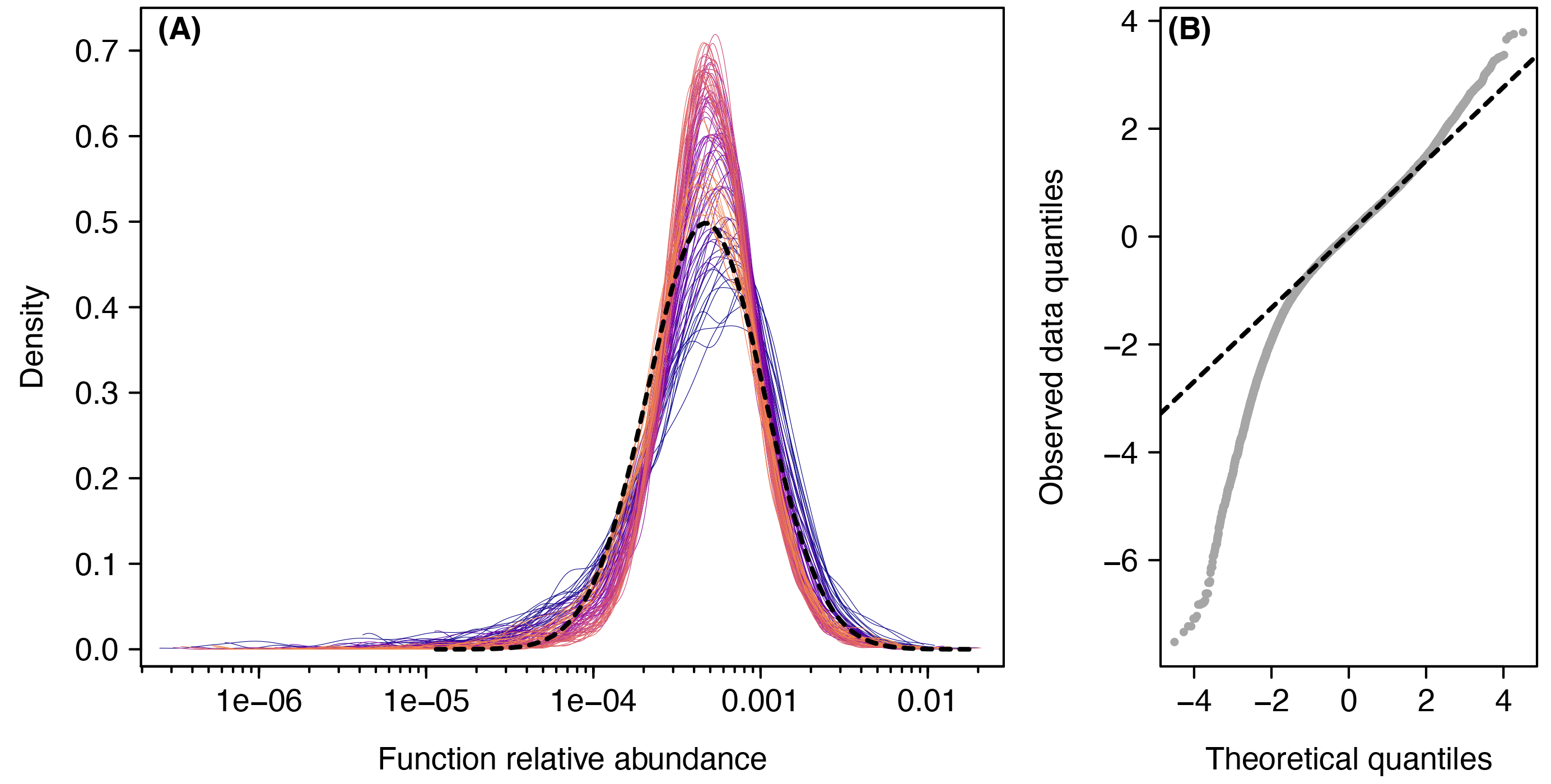}
	\caption{\textit{\textbf{(A)} Density of R function relative abundance across all GitHub repositories for each calendar month between January 2013 and December 2021. Density from a best-fitting normal distribution is overlaid (black dashed line), estimated via an intercept only regression fit with log-normal transformed relative abundances from all months. \textbf{(B)} Quantile-quantile plot of residuals from intercept-only regression with normal expectations overlaid (black dashed line).}}
	\label{fig:figS4}
\end{figure}

\begin{figure}[h]
	\centering
	\includegraphics[scale = 0.8]{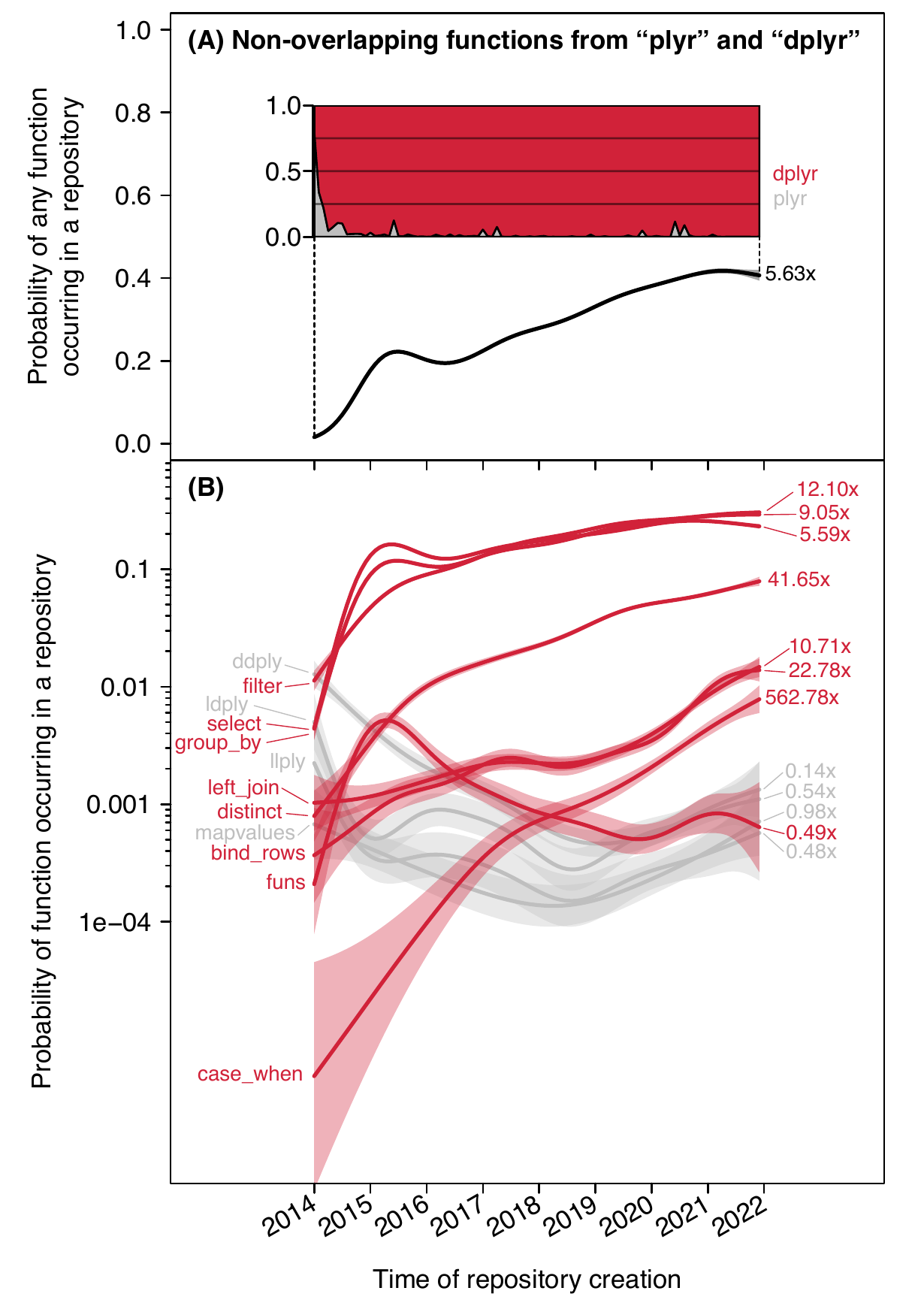}
	\caption{\textit{Occurrence probability patterns across \textbf{(A)} a selection of non-overlapping functions from the plyr and dplyr packages (the packages share developers and many function names). Y-axis values are the probability that a repository will use any function included in \textbf{(B)}. Functions are coloured based on package (dplyr, red; plyr, grey). Right-hand numbers are proportional changes from 2014 to 2021 yearly means (i.e. 2.0x = doubling in occurrence probability). Area plot in (A) is proportional breakdown of all function calls over time by package.}}
	\label{fig:figS5}
\end{figure}

\end{document}